\documentclass[aps,prl,letterpaper,10pt,twoside,tightenlines,nofootinbib,showpacs,preprint,twocolumn]{revtex4}
\usepackage{graphicx}
\usepackage[sort&compress]{natbib}
\usepackage{latexsym}
\usepackage{epsfig}
\usepackage{amsmath}

\usepackage{MnSymbol}

\def\be{\begin{equation}}
\def\ee{\end{equation}}

\begin{document}

\title{Universality in Hadronic and Nuclear Collisions  at High Energy}
\author{P. Castorina$^{1,2}$, A. Iorio$^{2}$, D. Lanteri$^{1,3}$, H. Satz$^4$ and M. Spousta$^{2}$}
\affiliation{
\mbox{${}^1$ INFN, Sezione di Catania, I-95123 Catania, Italy.}\\
\mbox{${}^2$ Institute of Particle and Nuclear Physics, Faculty of Mathematics and Physics, Charles University}\\
\mbox{V Hole\v{s}ovi\v{c}k\'ach 2, 18000 Prague 8, Czech Republic}\\
\mbox{${}^3$ Dipartimento di Fisica e Astronomia, Università  di Catania, I-95123 Catania, Italy.}\\
\mbox{${}^4$ Fakult\"at f\"ur Physik, Universit\"at Bielefeld, Germany.}
}

\date{\today}
\begin{abstract}
  Recent experimental results in proton-proton and in proton-nucleus  collisions at Large Hadron Collider energies
show a strong similarity to those observed in nucleus-nucleus  collisions, where the formation of a quark-gluon
plasma is expected. 
  We discuss the comparison between small colliding systems and nucleus-nucleus collisions, for: a)~the strangeness suppression factor $\gamma_s$ and yields of multi-strange hadrons; b)~the average transverse 
momentum, $p_t$, with particular attention to the low $p_t$ region where soft, non-perturbative effects are important; c)~the elliptic flow
scaled by the participant eccentricity.
The universal behavior in hadronic and nuclear high energy collisions emerges for all these observables in terms of  a specific dynamical variable which corresponds to 
  the entropy density of initial system in the collision  and which takes into account the transverse size of the initial configuration and its fluctuations.

\end{abstract}
 \maketitle

{\it Introduction:}
  Recent experimental results 
in proton-proton ($pp$) and proton-nucleus ($pA$) 
collisions at the Large Hadron Collider (LHC) and 
Relativistic Heavy Ion Collider (RHIC)
show a strong similarity to those observed in nucleus-nucleus ($AA$) collisions, where the formation of a quark-gluon plasma is expected.
Many different signatures~\cite{ALICE:2017jyt,Abelev:2013haa,Khachatryan:2016txc,atlas1,cms1,cms2,PHENIX:2018lia,p1,p2} support the conclusion that the system created in 
high energy, high multiplicity collisions with ``small'' initial settings, i.e. $pp$ and $pA$, is essentially the same as that one produced with ``large'' initial 
$AA$ configurations.

The ALICE collaboration reported~\cite{ALICE:2017jyt} the enhanced production of multi-strange hadrons, previously observed in $PbPb$ collisions \cite{ABELEV:2013zaa},
in high energy, high multiplicity, $pp$ events. The strangeness enhancement was suggested to be present in high-multiplicity $pp$ collisions on theoretical grounds in 
Refs.~\cite{cs1,cs2} by considering a specific dynamical variable corresponding to the initial entropy density of the collisions, which takes into account the 
transverse size (and its fluctuations) of the initial configuration in high multiplicity events~\cite{cs3,cs4}.
  Noticeably, the energy loss in $AA$ collisions was also shown to scale in the same dynamical variable~\cite{loss}.

  An important similarity between $pp$, $pA$, and $AA$ collisions was identified also in several measurements of long-range di-hadron azimuthal correlations 
\cite{Aad:2012gla,atlas1,cms2,Khachatryan:2016txc} indicating universally present flow-like patterns.

The similarity of the average transverse momentum ($p_t$) between $pp$, $pA$, and $AA$ collisions was discussed in Refs.~\cite{larry1,larry2,larry3} where the scaling 
of $p_t$ as a function of the variable $N_{track}/A_T$ ($N_{track}$ being the multiplicity and $A_T$ the transverse area of the initial system) was explored in the 
framework of Color Glass Condensate (CGC), where also the geometrical scaling of direct-photon production in hadron collisions at RHIC and LHC energies has been 
obtained in terms of the saturation scale, proportional to the transverse entropy density~\cite{larryphoton}.

\vskip 10pt

The previously discussed similarities indicate that a few dynamical ingredients, common to the different initial settings, drive 
the particle production, independently of the complexity of the non-equilibrium dynamics with annihilation/creation of many interacting quarks and gluons and 
hadronization of final partons.

  In this paper, we discuss some pieces of the mosaic of this universal behavior in a unified way. More precisely, for small colliding systems versus $AA$ collisions, we compare: a)~the 
strangeness enhancement, refining previous analyses~\cite{cs3,cs4}; b)~the mean $p_t$, with particular attention to the low $p_t$ region where the soft, 
non-perturbative effects are important; c)~the elliptic flow, $v_2$, where a scaling behavior has been already 
observed~\cite{Alver:2007rs,Chatrchyan:2012ta,Manly:2005zy} for different nuclei ($Au$, $Cu$ and $Pb$) by considering the ratio $v_2/\epsilon_{part}$, where 
$\epsilon_{part}$ is the participant eccentricity, defined for example in~\cite{Manly:2005zy,Bhalerao:2006tp,Alver:2008zza,Hirano:2009ah}.
  All the comparisons speak in favor of the universal, initial entropy density driven mechanism for the particle production across different colliding systems.

\vskip 10pt

{\it Emergent Universality: }
  Let us recall that the initial entropy density $s_0$ is given in the one-dimensional hydrodynamic
formulation~\cite{Bj} by the form
\begin{equation}
\begin{split}
s_0 ~\!\tau_0 &\simeq \frac{1.5}{A_T}\; \frac{dN^x_{ch}}{dy} =
\frac{1.5}{A_T}\;\frac{N_{part}^x}{2} \left. \frac{dN^x_{ch}}{dy}\right|_{y=0}\;,
\label{star0}
\end{split}
\end{equation}
with $x \simeq pp$, $pA$, $AA$.
Here $A_T$ is the transverse area, $(dN^x_{ch} / dy)_{y=0}$ denotes the number of produced charged secondaries,
normalized to half the number of participants $N_{part}^x$, in reaction 
$x$, and $\tau_0$ is the formation time.
  The initial entropy density is directly related to the number of partons per unit of transverse area and, 
due to the large fluctuations in high multiplicity events, one needs a reliable evaluation of the transverse area for different collisions.

In studying the strangeness enhancement and the average $p_t$, we use results from Glauber Monte Carlo (MC) ~\cite{Loizides:2017ack} to obtain $A_T$ as a function of 
multiplicity for $AA$ and for $pPb$ collisions. For $pp$ collisions the effective transverse area is sensitive to the fluctuations of the gluon field configurations 
and therefore we apply the CGC parameterization of the transverse size as a function of multiplicity~\cite{larry1,larry2,larry3}.

On the other hand, for the scaling behavior of the elliptic flow, namely of the ratio $v_2/\epsilon_{part}$,
the effective transverse area, $S$, of the initial setting is the one related to $\epsilon_{part}$. The $S$ is evaluated by MC simulations
in Refs.~\cite{Chatrchyan:2012ta,Hirano:2009ah} for $AA$ and in Ref.~\cite{Avsar:2010rf} for $pp$ collisions. 

\vskip 10pt

{\it Universality in strangeness production: }
  In Refs.~\cite{cs3,cs4} the parameter $\gamma_s \le 1$, which describes the strangeness suppression in the statistical hadronization model (SHM)~\cite{SHM}, was studied
as a function of the variable from Eq.~(1), by using an approximate evaluation of the transverse area for $pp$, $pPb$, and $AA$ collisions.

\begin{figure}	
	\includegraphics[width=\columnwidth]{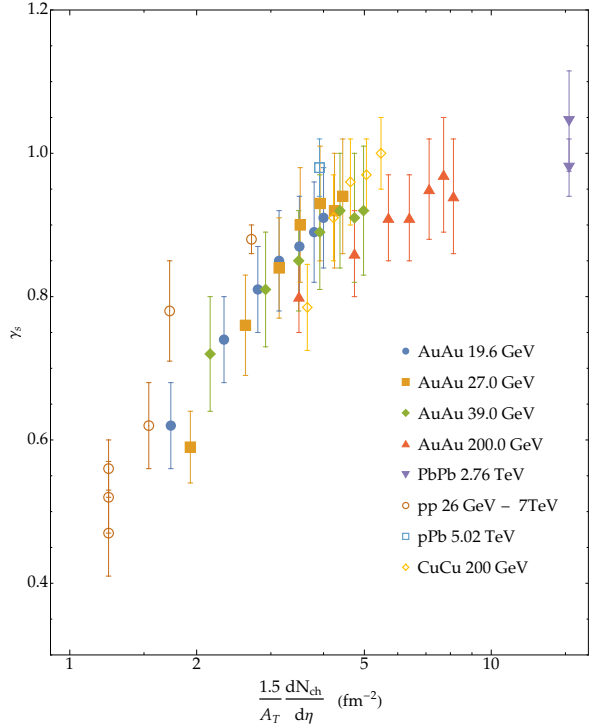}
	\caption{
          The strangeness suppression factor $\gamma_s$ as a function of initial entropy density evaluated for data from 
Refs.~\cite{Becattini:1996gy,Takahashi:2008yt,Becattini:2012xb,Adamczyk:2017iwn}. The Phobos parameterization~\cite{Alver:2010ck} for the relation between charge multiplicity, 
energy and the number of participants is applied for RHIC data.}
	\label{Fig:1}
\end{figure}

Here we use an improved evaluation of the transverse area for $AA$, $pPb$, and $pp$ collisions as described in the previous section. The resulting scaling behavior for 
the strangeness production is reported in Fig.~\ref{Fig:1}, where $\gamma_s$ for $AA$ at different energies and centralities are shown along with those for $pPb$ and $pp$ 
collisions. The data refers to $pp$ at energy $\sqrt{s}=26$~GeV$-7$~TeV~\cite{Becattini:1996gy,Takahashi:2008yt}, to $pPb$ at $\sqrt{s}=2.76$~TeV~\cite{io6,cs1,cs2,cs3,cs4}, to 
$PbPb$ at $\sqrt{s}=2.76$~TeV~\cite{Becattini:2012xb}, to $AuAu$ at $\sqrt{s}=19.6$, $27$, $39$ and $200$~GeV~\cite{Adamczyk:2017iwn}, and to $CuCu$ at 
$\sqrt{s}=200$~GeV~\cite{Takahashi:2008yt}.

\begin{figure}	
	\includegraphics[width=\columnwidth]{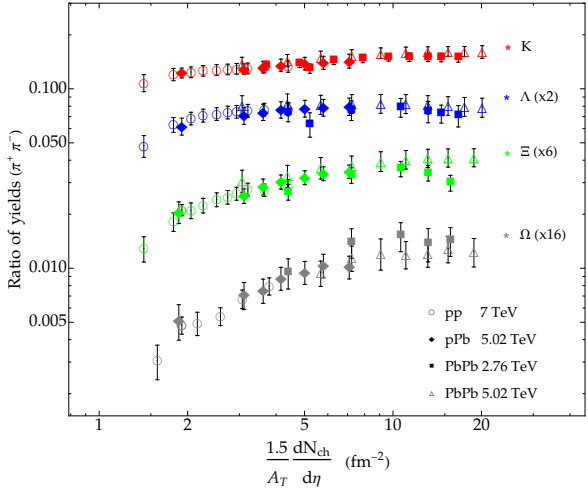}
	\caption{
          The strangeness production quantified in terms of the ratio of yields of K, $\Lambda$, $\Xi$, and $\Omega$ hadrons to pions evaluated as a function of initial 
entropy density for data from~Refs.~\cite{Abelev:2013vea,Abelev:2013xaa,Acharya:2019yoi,Kalinak:2017xll,Adam:2015vsf,Abelev:2013haa,ALICE:2017jyt,ABELEV:2013zaa}.} 
	\label{Fig:2}
\end{figure}

  The universal trend shows that $\gamma_s$ increases with the parton density in the transverse plane, up to the fixed point $\gamma_s=1$, where any suppression 
disappears.
  The universal trend of strangeness production versus entropy density in $PbPb$~\cite{Abelev:2013vea,Abelev:2013xaa,ABELEV:2013zaa,Kalinak:2017xll}, 
$pPb$~\cite{Adam:2015vsf,Abelev:2013haa}, $pp$~\cite{ALICE:2017jyt,Acharya:2019yoi} is shown in Fig.~\ref{Fig:2} for different particle species.

\vskip 10pt

\begin{figure}	
	\includegraphics[width=\columnwidth]{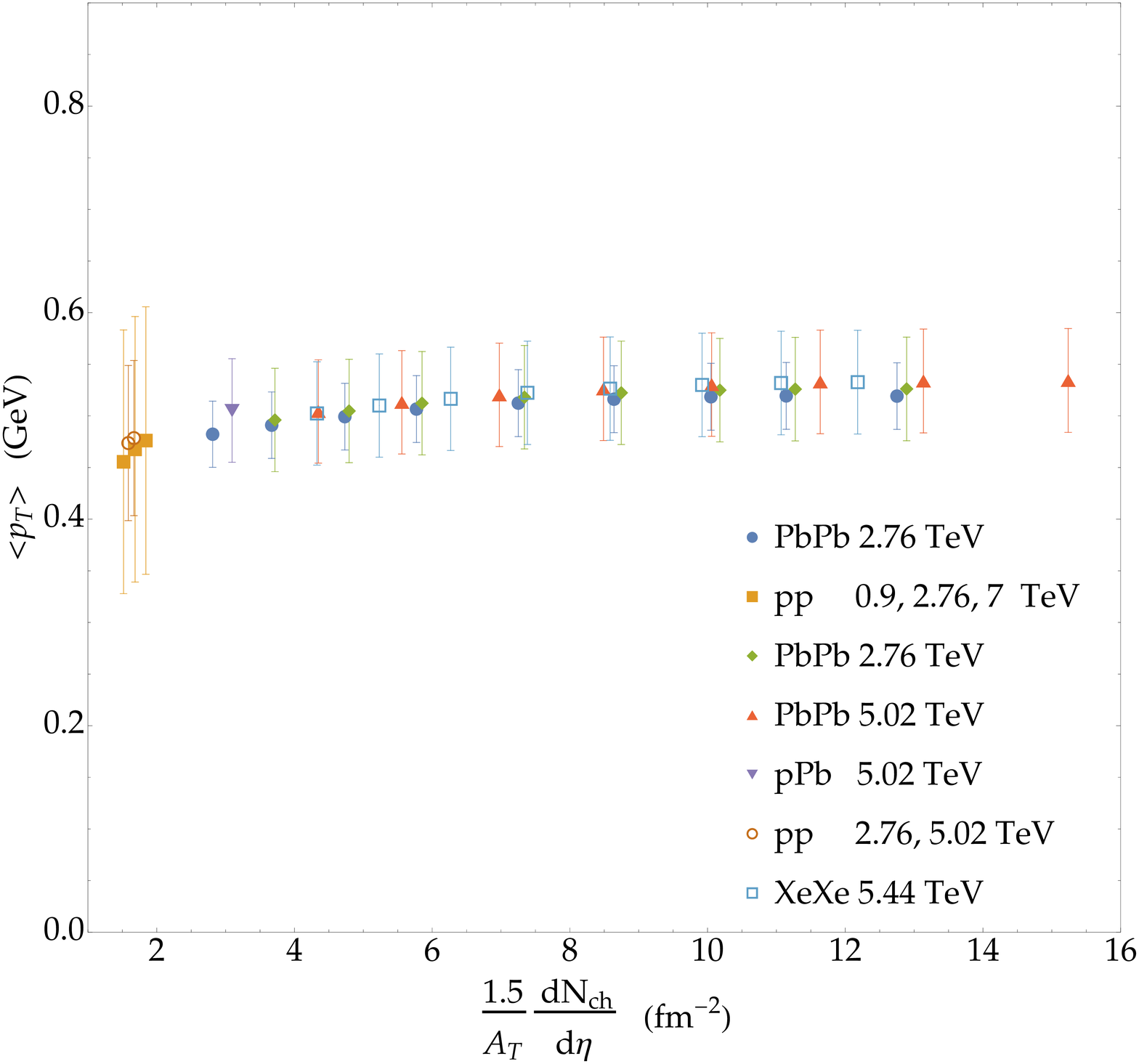}
	\caption{Average $p_t$ as a function of initial entropy density evaluated in the interval of $0.15 < p_t < 1.15$~GeV for the data from 
Refs.~\cite{Abelev:2012hxa,Abelev:2013ala,Acharya:2018qsh,Acharya:2018eaq}.}
	\label{fig:3}
\end{figure}

{\it Universality in mean transverse momentum: }
  As recalled, the scaling of the average $p_t$ as a function of the  variable $N_{track}/A_T$  has been discussed  in~\cite{larry1,larry2,larry3}.
  Here we analyze the average $p_t$ in the low transverse momentum region where the soft, non-perturbative effects in the particle production are more important due to running of the strong coupling constant than in the higher $p_t$ range. 
  The behavior of the average $p_t$ is evaluated in the region $0.15 < p_t < 1.15$~GeV for different colliding systems as a function of the dynamical variable from
Eq.~\eqref{star0}.
  The results are shown in Fig.~\ref{fig:3} for the data from Refs.~\cite{Abelev:2012hxa,Abelev:2013ala,Acharya:2018qsh,Acharya:2018eaq}.
  One can see that the average $p_t$ for soft particle production follows the same slowly increasing trend for all the collisional systems.
  Equally good scaling was obtained also for the extended region $0.15 < p_t < 1.5$~GeV for which only a subset of data can be used to evaluate average-$p_t$ due to 
differences in the binning across experiments.

\vskip 10pt

{\it Universality in the elliptic flow: }
  In non-central collisions, the beam direction and the impact parameter vector define a reaction plane for each event. If the nucleon density within the nuclei is continuous, the initial
nuclear overlap region has an ``almond-like'' shape and the impact parameter determines uniquely the initial geometry of the collision. 

In a more realistic description, where the position of the individual nucleons that participate in inelastic interactions is considered, the overlap region has a more 
irregular shape and the event-by-event orientation of the almond fluctuates around the reaction plane \cite{Alver:2006wh,Alver:2008zza}. Therefore, in the analysis of 
the elliptic flow where the fluctuations are important, the geometrical eccentricity is replaced by the participant eccentricity, $\epsilon_{part}$, defined using the 
actual distribution of participants. The size of the fluctuation in $\epsilon_{part}$ and its correlated transverse area $S$ (different from the geometrical one) are 
evaluated by Glauber MC as previously described.

\begin{figure}	
	\includegraphics[width=\columnwidth]{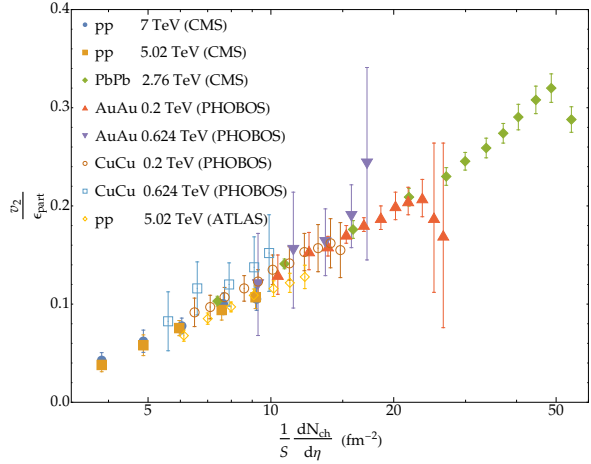}
	\caption{
        The $v_2/\epsilon_{part}$ values for $pp$, $PbPb$, $AuAu$, and $CuCu$ evaluated as a function of entropy density for data 
from Refs.~\cite{Alver:2007rs,Chatrchyan:2012ta,Khachatryan:2016txc,Aaboud:2016yar}.}
        \label{fig:4}
\end{figure}

The scaling of $v_2/\epsilon_{part}$ versus the initial entropy density is depicted in Fig.~\ref{fig:4} for $AA$ ~\cite{Alver:2007rs,Chatrchyan:2012ta} and $pp$~\cite{Khachatryan:2016txc,Aaboud:2016yar}.
One can see that the $pp$ trend, at lower values, is smoothly followed by the data-points from $AA$ collisions.

\vskip 10pt

{\it How to check the universal trend:}
  The analyses of $\gamma_s$, average $p_{T}$, and $v_2/\epsilon_{part}$ presented above support the conclusion that at fixed entropy density the 
``coarse-grain'' features of the quark-gluon system formed in high energy collisions are independent of the initial configuration. The scaling variable 
(Eq.~\eqref{star0}) is a function of multiplicity and the transverse area, and one can evaluate at which multiplicity one can expect the same behavior in 
high-multiplicity $pp$ and $PbPb$ collisions, by solving the equation $(dN/d\eta)_{AA}/A_{T}^{AA} = x/A_T^{pp}(x)$ for $x$ being the multiplicity in $pp$. The result is 
shown in Tab.~\ref{tab:2} for $PbPb$ collisions at $5.02$ TeV which represent the largest available heavy-ion dataset at the LHC.
  The values from Tab.~\ref{tab:2} can be used in subsequent experimental or phenomenological studies aiming to further check the universal trends 
in hadronic and nuclear collisions using high-multiplicity $pp$ collisions at the largest available LHC energies.

\begin{table}
	\begin{tabular}{|c|c|c|c|}
		\hline
		$\;\displaystyle \frac{1.5}{A_T}\;\frac{dN_{ch}}{d\eta}\;$& $\displaystyle\left(\frac{dN_{ch}}{d\eta}\right)_{pp}$&$\displaystyle \left(\frac{dN_{ch}}{d\eta}\right)_{PbPb}$&  \begin{tabular}{c}Pb-Pb\\Centrality\end{tabular} \\
		\hline 
		$20.1\pm 0.8$ & $100.\pm 4.$ &$ 1943.\pm 56.$ & \text{0-5$\%$} \\
		$17.5\pm 1.1$ & $87.\pm 5.$ & $1587.\pm 47. $& \text{5-10$\%$} \\
		$15.4\pm 0.9$ & $76.\pm 4.$ & $1180.\pm 31. $& \text{10-20$\%$} \\
		$12.2\pm 0.6$ & $60.6\pm 3.1$ &$ 649.\pm 13.$ & \text{20-40$\%$} \\
		$8.3\pm 0.7$ & $41.2\pm 3.4$ & $251.\pm 7. $& \text{40-60$\%$} \\
		$5.2\pm 0.8$ & $26.\pm 4.$ & $70.6\pm 3.4 $& \text{60-80$\%$} \\
		$3.1\pm 1.1$ & $12.4\pm 3.0$ & $17.5\pm 1.8$ & \text{80-90$\%$} \\
		\hline 
	\end{tabular}
	\caption{$dN_{ch}/d\eta$ in $PbPb$ at $5.02$~TeV and $pp$ for different values of the variable in Eq.~\eqref{star0}.}
	\label{tab:2}
\end{table}

\vskip 10pt

{\it Comments and Conclusions: }
 High energy, high multiplicity events produced in small colliding systems show dynamical behavior very similar to that present in $AA$ collisions. Observations made in 
this paper suggest that the dynamical behavior is largely driven by the initial entropy, density that is by the parton density in the transverse plane. A clear 
quantification of limits on the presence of jet quenching in small colliding systems (see e.g. discussions and new measurements in Refs.~\cite{mangano,p5,Aad:2019ajj}) 
or more detailed correlation measurements (see e.g. recent work in Refs.~\cite{Acharya:2019vdf,Sirunyan:2017igb,Aad:2019fgl}) may help to improve understanding of this 
similarity. This kind of measurements can be done in details at the LHC or RHIC or at a 100 TeV $pp$ collider which is considered for the future \cite{mangano2} and 
which would significantly enhance the reach of $pp$ collisions in multiplicities.

\vskip 10pt

{\it Acknowledgments: } M.S. is supported by grants GA\v CR 18-12859Y and UNCE/SCI/013.

\end{document}